\documentclass{elsart}

\usepackage{graphicx}

\usepackage{amssymb}

\begin{document}
\runauthor{Amritkar and Jalan}
\begin{frontmatter}
\title{Coupled Dynamics on Networks}
\author{R. E. Amritkar \thanksref{amritkar}}
\thanks[amritkar]{amritkar@prl.res.in}
\author{and Sarika Jalan\thanksref{sarika}}
\thanks[sarika]{sarika@prl.res.in}

\address{Physical Research Laboratory, Navrangpura,
Ahmedabad-380009, India}

\begin{abstract}
We study the synchronization of coupled dynamical systems
on a variety of networks. 
The dynamics is governed by a local nonlinear 
map or flow for each node of the network and couplings connecting different 
nodes via the links of the network. For small coupling strengths nodes
show turbulent behavior but form synchronized clusters as
coupling increases. When nodes show synchronized behaviour, we observe 
two interesting phenomena. First,
 there are some nodes
of the floating type that show intermittent behaviour between getting
attached to some clusters and evolving independently. Secondly,
we identify two
different ways of cluster formation, namely self-organized clusters 
which have mostly
intra-cluster couplings and driven clusters which have mostly 
inter-cluster couplings. 
\end{abstract}
\end{frontmatter}

\section{Introduction}

Several complex systems have underlying structures that are 
described by networks or graphs
\cite{Strogatz,rev-Barabasi}. Recent interest in networks is due to
the discovery that several naturally occurring
networks come under some universal classes and they can
be simulated with simple mathematical models, viz small-world 
networks \cite{Watts}, scale-free networks \cite{scalefree} etc. 

Several networks in the real world consist of dynamical elements interacting
with each other and the interactions define the links of the network.
Several of these networks have a large number
of degrees of freedom and it is important to understand their dynamical
behavior. Here, we study the synchronization
and cluster formation in networks consisting of interacting
dynamical elements. A general model of coupled dynamical systems on
networks will consist
of the following three elements.
\begin{enumerate}
\item The evolution of uncoupled elements.
\item The nature of couplings.
\item The topology of the network.
\end{enumerate}  

Most of the earlier studies of synchronized cluster formation in
coupled chaotic systems have focused on
networks with large number of connections ($\sim N^2$)
\cite{rev-Kaneko}. 
 In this paper, we consider networks
with number of connections
of the order of $N$. This small number of connections
allows us to study the role that different connections play in
synchronizing different nodes and
the mechanism of synchronized cluster formation. The study reveals two 
interesting phenomena. First,
when nodes form synchronized clusters, there can be some nodes which
show an intermittent behaviour between independent evolution and
evolution synchronized with some cluster. Secondly,  
the cluster formation can be in two different ways, driven and self-organized
phase synchronization \cite{sarika-REA1}.
The connections or couplings in the self-organized phase synchronized
clusters are mostly of the intra-cluster type while those in the
driven phase synchronized clusters are mostly of
the inter-cluster type.

\section{Coupled dynamical systems and synchronized clusters}

Consider a network of $N$ nodes and $N_c$ connections (or couplings)
between the nodes. Let each node of the network be assigned an $m$-dimensional
dynamical variable ${\bf x}^i, i=1,2,\ldots,N$. A very general dynamical
evolution can be written as
\begin{equation}
\frac{d{\bf x}_i}{dt} = {\bf F}(\{ {\bf x}_i \}).
\end{equation}
In this paper, we consider a separable case and the evolution equation 
can be written as,
\begin{equation}
\frac{d{\bf x}_i}{dt} = {\bf f}({\bf x}_i) + \frac{\epsilon}{k_i}
\sum_{j \in \{k_i\}} {\bf g}({\bf x}_j).
\label{evol-cont}
\end{equation}
where $\epsilon$ is the coupling constant, $k_i$ is the degree of node
$i$, and $\{k_i\}$ is the set of nodes connected to the node $i$.
A sort of diffusion version of the evolution equation~(\ref{evol-cont}) is
\begin{equation}
\frac{d{\bf x}_i}{dt} = {\bf f}({\bf x}_i) + \frac{\epsilon}{k_i}
\sum_{j \in \{k_i\}} \left({\bf g}({\bf x}_j)-{\bf g}({\bf
x}_i)\right).
\label{evol-cont-diff}
\end{equation} 

Discrete versions of Eqs.~(\ref{evol-cont})
and~(\ref{evol-cont-diff}) are
\begin{equation}
{\bf x}_i(t+1) = {\bf f}({\bf x}_i(t)) + \frac{\epsilon}{k_i}
\sum_{j \in \{k_i\}} {\bf g}({\bf x}_j(t)).
\label{evol-disc}
\end{equation}
and
\begin{equation}
{\bf x}_i(t+1) = {\bf f}({\bf x}_i(t)) + \frac{\epsilon}{k_i}
\sum_{j \in \{k_i\}} \left({\bf g}({\bf x}_j(t))-{\bf g}({\bf
x}_i(t))\right).
\label{evol-disc-diff}
\end{equation}  

For the discrete evolution we use logistic or circle maps while for
the continuous case we use Lorenz or R\"ossler systems.

\subsection{Phase synchronization and synchronized clusters}

Synchronization of coupled dynamical systems 
\cite{book-syn} is manifested by the appearance of some
relation between the functionals of different dynamical variables. 
The exact synchronization corresponds to
the situation where the dynamical variables for different nodes have identical
values. The phase synchronization corresponds to the situation where the 
dynamical variables 
for different nodes have some definite relation between their phases 
\cite{phase1,phase2}. When the number 
of connections in the network
is small ($N_C \sim N$) and when the local dynamics of the
nodes (i.e. function $f(x)$) is in the chaotic zone, and we look at
exact synchronization, we find that only
few synchronized clusters with small number of nodes are formed.
However, when we look at phase synchronization, synchronized clusters 
with larger number of nodes are obtained. Hence, in our numerical study we
concentrate on phase synchronization. 

\section{General properties of synchronized dynamics}

We consider some general properties of synchronized dynamics. 
They are valid
for any coupled discrete and continuous dynamical systems.
Also, these
properties are applicable for exact as well as phase or any other type 
of synchronization
and are independent of the type of network. 

\subsection{Behavior of individual nodes}
As the network evolves, it splits into several synchronized clusters. 
Depending on their asymptotic dynamical behaviour the nodes
of the network can be divided into three types. \\
(a) {\it Cluster nodes}: A node of this type synchronizes with other nodes and 
forms a synchronized cluster. Once this node enters a synchronized cluster
it remains in that cluster afterwards. \\
(b) {\it Isolated nodes}: A node of this type does not synchronize
with any other node 
and remains isolated for all the time. \\
(c) {\it Floating Nodes}:  A node of this type keeps on switching intermittently
between an independent evolution and a synchronized evolution
attached to some cluster.

Of particular interest are the floating nodes and we will discuss some
of their properties afterwards.

\subsection{Mechanism of cluster formation} 
The study of the relation between the synchronized clusters and the couplings
between the nodes represented by the
adjacency matrix $C$ shows two different 
mechanisms of cluster formation \cite{sarika-REA1,pre2}. \\
(i) Self-organized clusters: The nodes of a cluster can be 
synchronized because of intra-cluster
couplings. We refer to this as
the self-organized synchronization and 
the corresponding synchronized clusters as self-organized clusters. \\
(ii) Driven clusters: The nodes of a cluster can be
synchronized because
of inter-cluster couplings. 
Here the nodes of one cluster are driven by those
of the others. We refer to this as the driven
synchronization and the corresponding clusters as driven clusters.

In our numerical studies we have been able to identify ideal clusters
of both the types,
as well as clusters of the mixed type where both ways of
synchronization contribute
to cluster formation. 
(Fig.~1 of Ref.~\cite{sarika-REA1} gives examples of ideal as well as
mixed clusters in coupled map networks.) 
In general we find
that the scale free
networks and the Caley tree networks lead to better cluster formation than 
the other types of networks with the same average connectivity. 

\begin{figure}[hb]
\begin{center}
\includegraphics[width=14cm]{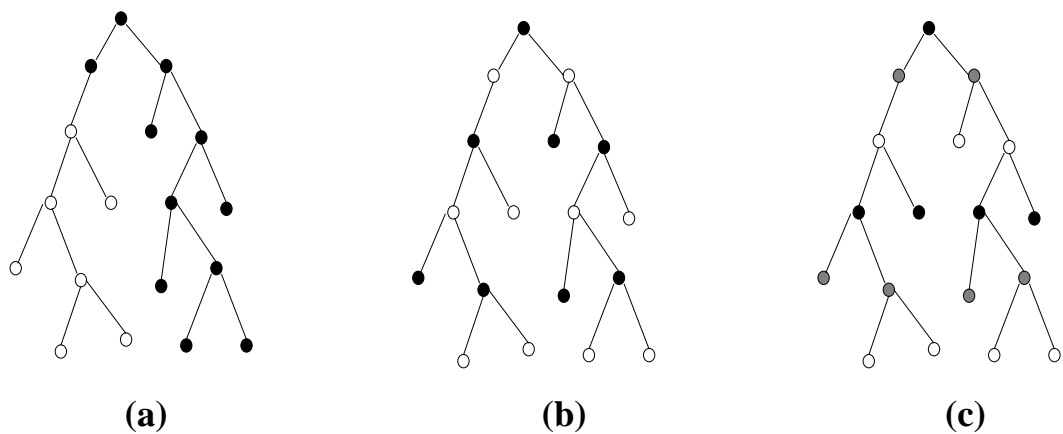}
\end{center}
\caption{Different ways of cluster formation in a tree structure are
demonstrated. The open, solid and gray circles show nodes belonging to 
different clusters. (a) shows two clusters of the self-organized
type, (b) shows two clusters of driven type and (c) shows three
clusters of the driven type.}
\label{tree-clus}
\end{figure}

Geometrically the two mechanisms of cluster formation can be easily
understood by considering a tree type network. A tree
can be broken into different clusters in different ways. \\
(a) A tree can be broken into two or more disjoint clusters with only
intra-cluster couplings by breaking one
or more connections. Clearly, this splitting is not unique
and will lead to self-organized clusters. Figure~\ref{tree-clus}(a)
shows a tree forming two synchronized clusters of self-organized
type. This situation is
similar to an Ising ferromagnet where domains of up and down spins can 
be formed. \\
(b) A tree can 
also be divided into two clusters by putting connected nodes into different
clusters. This division is unique and leads to two clusters with only
inter-cluster couplings, i.e. driven clusters. Figure~\ref{tree-clus}(b)
shows a tree forming two synchronized clusters of the driven type. 
This situation is
similar to an Ising anti-ferromagnet where two sub-lattices of up and 
down spins are formed.\\
(c) Several other ways of splitting a tree are possible. E.g. it is easy
to see that a tree can be broken into three clusters of the driven
type. This is shown in figure~\ref{tree-clus}(c). There is no simple
magnetic analog for this type of cluster formation. It
can be observed close to a period
three orbit. We note that four or more clusters of the driven type are 
also possible. As compared to the cases (a) and (b) discussed above
which are commonly observed, 
the clusters of case (c) are not so common and are
observed only for some values of the parameters. 

\section{Linear stability analysis}

A suitable network to study the stability of self-organized
synchronized clusters is the globally coupled network. The stability
of globally coupled maps is well studied in the literature
\cite{GCM-stab1,GCM-stab2,GCM-stab3}.
An ideal example to consider the stability of the driven synchronized
state is a complete bipartite network.
A complete bipartite network consists of two sets of nodes with each
node of one set connected
with all the nodes of the other set and no connection between the
nodes of the same set. Let $N_1$ and $N_2$ be the number of nodes 
belonging to
the two sets. We define a bipartite synchronized
state as the one that 
has all $N_1$ elements of the first set synchronized
to some value, say ${\bf X}_1(t)$, and 
all $N_2$ elements of the second set  synchronized
to some other value, say ${\bf X}_2(t)$.

 All the eigenvectors and the eigenvalues of the Jacobian matrix for the bipartite synchronized state can be determined explicitly.
The eigenvectors of the type
$(\alpha,\ldots,\alpha,\beta,\ldots,\beta)^T$ determine the
synchronization manifold and this manifold has dimension two. 
All other eigenvectors correspond to
the transverse manifold.
Lyapunov exponents corresponding to the transverse eigenvectors
for Eq.~(\ref{evol-disc-diff}) with one dimensional variables and $g(x)=f(x)$
are
\begin{eqnarray}
\lambda_1 &=& \ln|(1-\epsilon)| + \frac{1}{\tau} \lim_{\tau \to\infty}
\sum^\tau_{t=1} \ln |f^{\prime}(X_1)|, 
\nonumber \\
\lambda_2 &=& \ln|(1-\epsilon)| + \frac{1}{\tau} \lim_{\tau \to\infty}
\sum^\tau_{t=1} \ln |f^{\prime}(X_2)|,
\label{lya-driven-Nlarge}
\end{eqnarray}
and $\lambda_1$ and $\lambda_2$ are respectively $N_1-1$ and $N_2-1$
fold degenerate \cite{pre2}. Here, $f^{\prime}_(X_1)$ and $f^{\prime}_(X_2)$ are
the derivatives of $f(x)$  at $X_1$ and $X_2$ respectively.
The synchronized state is stable provided the transverse Lyapunov
exponents are negative.
If $f^\prime$ is bounded then from
Eqs.~(\ref{lya-driven-Nlarge}) we see that for $\epsilon$ larger than
some critical value, $\epsilon_b (<1)$, bipartite synchronized state will be
stable. Note that this bipartite synchronized state will be stable even if one
or both the remaining Lyapunov exponents corresponding to the
synchronization manifold are
positive, i.e. the trajectories are chaotic. The linear stability 
analysis for other type of couplings and dynamical systems can be done 
along similar lines.

\section{Floating nodes}
We had noted earlier that the nodes can be divided into three types,
namely cluster nodes, isolated nodes and floating nodes, depending on
the asymptotic behavior of the nodes. 
Here, we discuss some properties of the floating nodes which show an
intermittent behavior between synchronized evolution with some cluster 
and an independent evolution.

\begin{figure}
\begin{center}
\includegraphics[width=10cm]{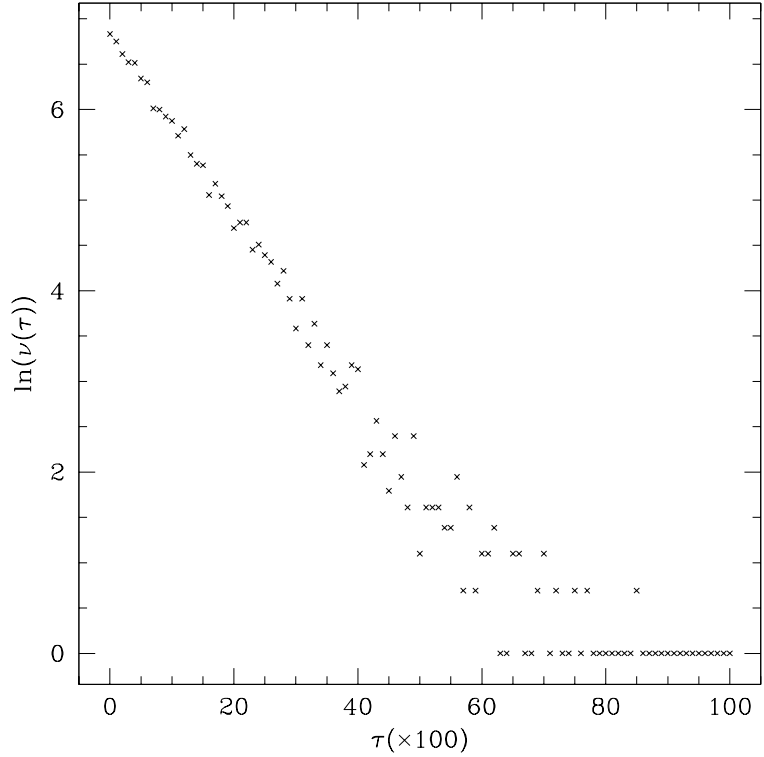}
\end{center}
\caption{The figure plots the frequency of residence time $\nu(\tau)$
of a floating node in a cluster as a
function of the residence time $\tau$. A good straight line fit on log-linear
plot shows exponential dependence.}
\label{freq-floating}
\end{figure}

Let $\tau$ denote the residence time of a floating node in a cluster 
(i.e. the continuous time interval that the node is in a cluster).
Figure~\ref{freq-floating} plots the frequency of residence time $\nu(\tau)$
of a floating node as a
function of the residence time $\tau$. 
A good straight line fit on log-linear
plot shows an exponential dependence, 
\begin{equation}
\nu(\tau) \sim \exp(-\tau/ \tau_r)
\label{exp-dist-floating}
\end{equation} 
where
$\tau_r$ is the typical residence time
for a given node. We have also studied the distribution of the time
intervals for which a floating node is not synchronized with a given
cluster. This also shows an exponential distribution.  

Let us now consider the condition for the occurrence of floating nodes.
Consider a floating node in a cluster. The stability of this cluster
is ensured if the transverse Lyapunov exponents are all negative. The
floating node will leave the cluster
provided the conditional Lyapunov exponent for this node,
assuming that the other nodes in the cluster remain synchronized,
changes sign and becomes positive. Thus the fluctuation of the
conditional Lyapunov exponent about zero can be taken as the condition 
for the existence of a floating node.

Several natural systems show examples of floating nodes, 
e.g. some birds may show intermittent behaviour between free flying and
flying in a flock. An interesting example in physics is that of particles
or molecules in a liquid in
equilibrium with its vapor where the particles
intermittently belong to the liquid and vapor. Under suitable
conditions it is possible to 
argue that the residence time of a tagged particle in the liquid phase 
should have an exponential distribution \cite{pre2}, i.e. a behavior similar to
that of the floating nodes (Eq.~(\ref{exp-dist-floating})).

\section{Conclusion and Discussion}
We have studied the properties of coupled dynamical elements on
different types of networks. We find that
in the course of time evolution they form synchronized 
clusters. 

In several cases when synchronized clusters are formed there are some
isolated nodes which do not belong to any cluster. More interestingly
there are some {\it floating} nodes which show an intermittent behavior
between an independent evolution and an evolution synchronized with some
cluster. The residence time spent by a floating node in the synchronized cluster 
shows an exponential distribution.

We have identified two mechanisms of cluster formation, 
self-organized and driven phase synchronization. For self-organized
clusters intra-cluster couplings dominate while for driven clusters
inter-cluster couplings dominate.


\begin{thebibliography}{99}
\bibitem{Strogatz} S. H. Strogatz, {\em Nature\/} {\bf 410} (2001) 268 and
references theirin.
\bibitem{rev-Barabasi} R. Albert and A. L. Barab\"asi,
{\em Rev. Mod. Phys.\/} {\bf 74} (2002) 47 and references theirin.
\bibitem{Watts} D. J. Watts and S. H. Strogatz, {\em Nature (London)\/} {\bf 393}
(1998) 440.
\bibitem{scalefree} A. -L. Barab\"asi, R. Albert, {\em Science\/} {\bf
286} (1999) 509.
\bibitem{sarika-REA1} S. Jalan, R. E. Amritkar, {\em Phys. Rev. Lett.\/} 
{\bf 90} (2003) 014101.

\bibitem{rev-Kaneko} K. Kaneko, {\em Physica\/} {\bf D124} (1998) 
322.

\bibitem{book-syn} A. Pikovsky, M. Rosenblum and J. Kurth, 
{\em Synchronization : A universal concept in nonlinear dynamics}, 
Cambridge University Press, 2001. 
 
\bibitem{phase1} M. G. Rosenblum, A. S. Pikovsky, and J.
Kurth, {\em Phys. Rev. Lett.\/} {\bf 76}, (1996) 1804; W. Wang,
Z. Liu, Bambi Hu, {Phys. rev. Lett.\/} {\bf 84} (2000) 2610. 
\bibitem{phase2} S. C. Manrubia and A. S. Mikhailov, {\em Europhys. Lett.\/} 
{\bf 53} (4) (2001) 451.

\bibitem{pre2} S. Jalan, R. E. Amritkar and C. K. Hu, unpublished.

\bibitem{GCM-stab1} H. Fujisaka and T. Yamada, {\em Prog. Theo. Phys.\/} {\bf
69}, 32 (1983).
\bibitem{GCM-stab2} P. M. Gade, {\em Phys. Rev.\/} {\bf E54} (1996) 64.
\bibitem{GCM-stab3} M. Ding and W. Yang, {\em Phys Rev\/} {\bf E56}
(1997) 4009.

\end{thebibliography}
\end{document}